# Computational Orthodontic Force Simulation: A Review

Waheed Ahmad[1], Jing Xiong[1*], and Zeyang Xia[2*]

[1] Shenzhen Institutes of Advanced Technology, Chinese Academy of Sciences, Shenzhen, China

[2] School of Mechanical Engineering, Shanghai Jiao Tong University, Shanghai, China

**\*** Corresponding authors: jing.xiong@siat.ac.cn, zxia@sjtu.edu.cn

**Abstract**

In orthodontic treatment, the biological response of the tooth, periodontal ligament, and bone complex to orthodontic force is crucial in influencing treatment outcomes. The challenge lies in accurately measuring, estimating, and predicting these forces during clinical procedures. This review aims to fill the gap in the literature by systematically summarizing existing research on orthodontic force simulation, examining common loading techniques and technologies, and discussing the potential for refining the orthodontic force simulation process. The literature was comprehensively reviewed, with an emphasis on the exploration of the biological mechanism of tooth movement. Studies were categorized based on force-loading techniques for both fixed and invisible orthodontic appliances. Finite element (FE) analysis stands out as the predominant technique for orthodontic force simulation, with a significant focus on fixed orthodontics but limited emphasis on invisible orthodontics. Current orthodontic force simulations tend to be fragmented, often considering only the instantaneous response to applied forces. There exists an urgent demand for a sophisticated analytical simulation model. Such a model, possibly leveraging advanced technologies like deep learning, holds the promise of forecasting orthodontic treatment outcomes with heightened precision and efficiency.

## 1. Introduction

Orthodontics is a specialized branch of dentistry that treats malocclusion, establishes an equilibrium between adjacent soft tissues and the skeletal structures surrounding them, pursues aesthetic harmony, and prevents speech disorders and periodontal diseases [1]. Orthodontists use force as a therapeutic intervention to move specific teeth through the bone into their proper positions to achieve an optimal bite [2]. This process is a synergistic interplay of physical phenomena and biological tissue remodeling within the dentoalveolar complex.

The biological intricacies of orthodontic tooth movement involve the use of mechanical forces on teeth through fixed or removable devices. As force is applied, a series of physiological and biological reactions occur, leading to the vital remodeling of the alveolar bone [3]. Specifically, this force translates from the tooth to the alveolar bone, causing temporary changes to the PDL. Regions of bone remodeling and resorption are evident, often corresponding to areas of compression and tension around the tooth, based on the main forces at play (see figure 1). Moreover, the direction in which force is applied causes PDL compression on one side of the tooth, and on the other side, tension as the tooth moves towards the compressing bone [4]. This dynamic interplay between compression and tension and their effects on bone remodeling are essential for understanding tooth movement mechanics.

Further deepening the biological perspective, when pressure is applied on one side of a tooth, it releases the PDL fibers connecting the tooth to the bone. Conversely, movement of the dental root in the opposite direction causes tension in PDL fibers attached to the bone. This balance between compression and tension at the cellular level is crucial for bone remodeling, with bone loss on the compression side and bone formation on the tension side [4].

The accurate prediction of orthodontic forces is a significant clinical challenge, as these forces play a crucial role in influencing treatment outcomes. Despite detailed planning, the unpredictability in tooth movement necessitates sophisticated methods for precise force estimation and prediction during clinical procedures, emphasizing the need for advanced simulation techniques [5].

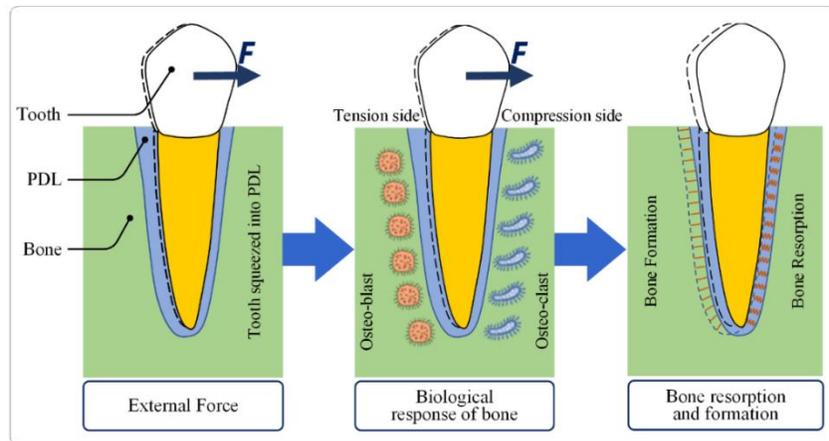

Figure 1. Illustration of the physiological and biomechanical responses in the periodontal ligament and alveolar bone during orthodontic force application, showcasing areas of compression and tension critical for bone remodeling.

From a simulation standpoint, understanding how these forces work and predicting their effects is pivotal. FE analysis emerges as a foremost method in capturing the transfer of forces from the tooth crown to the supporting structures of the alveolar bone via the PDL [6]. Since 1973, FE methods have been instrumental in quantifying the load on the periodontium non-destructively [7, 8]. But the creation of these biomechanical models requires a careful consideration of morphology, material properties, loads, and boundary conditions, presenting its own set of challenges [9, 10]. Importantly, the PDL is complex, and existing models often don't fully capture its detailed morphology, which can impact simulation results [7, 11]. The FE model building process, while varying in complexity, usually involves generating 3D models using CT and CBCT images, importing them into simulation environments, setting conditions, and analyzing results (see Figure 2).

Recent advancements in computational technologies, particularly deep learning, have opened new avenues in orthodontic simulations. These technologies promise to revolutionize the field by enhancing the accuracy of biomechanical models and forecasting orthodontic treatment outcomes with heightened precision. For instance, the use of deep learning in predicting post-orthodontic facial changes demonstrates the potential of these technologies in accurately forecasting treatment outcomes [12]. Additionally, the application of deep learning in segmenting CBCT images for orthodontic purposes highlights its role in improving the precision of biomechanical models [13]. This paper aims to explore these advancements and their potential impact on the future of orthodontic force simulations.

In the past, orthodontics has primarily depended on clinical experience for planning and predicting tooth movement. However, the intricate and sometimes unpredictable nature of orthodontic procedures, especially when using fixed appliances like bands, wires, and brackets [14], or invisible ones such as aligners [15], underscores the need for more precise predictions [16]. Given this context, the utilization of comprehensive orthodontic force simulations, especially FE analysis, has grown in importance for achieving more predictable outcomes. This paper aims to provide a thorough overview of the existing literature in the field of orthodontic force simulations. It will highlight common loading techniques, identify limitations in current studies, and advocate for a holistic simulation approach to improve the accuracy of tooth movement predictions.

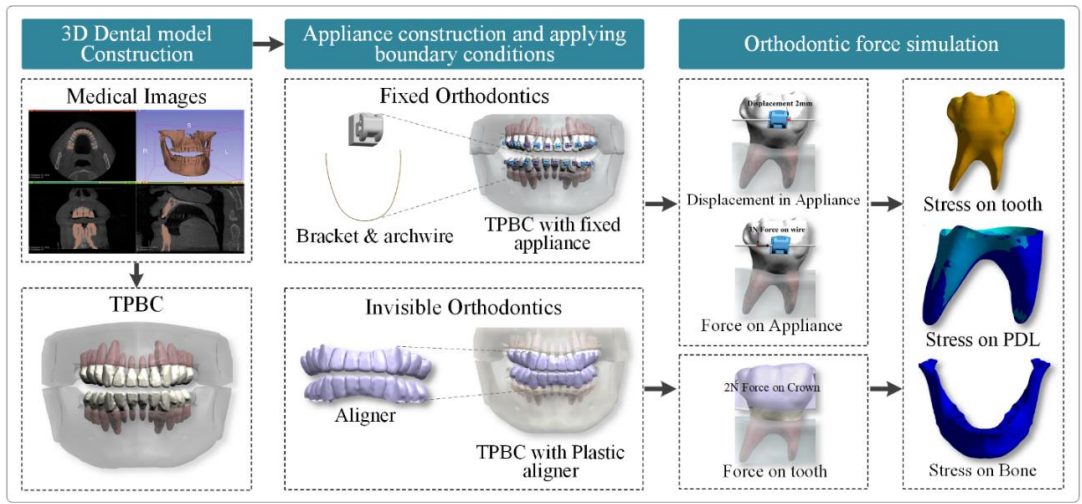

Figure 2. A comprehensive depiction of the sequential steps in orthodontic force simulation, from 3D modeling using CT/CBCT scans to the analysis of biomechanical effects in the dentoalveolar complex.

## 2. Literature Review Methods and Classification

### A. Selection Criteria

In our systematic review, we refined our selection criteria to focus on high-quality, clinically relevant studies in orthodontics and finite element analysis from 2005 to 2024, using databases like Google Scholar, Web of Science, Scopus, and PubMed. Keywords included "orthodontics," "force," "simulation," and "finite element analysis." Initially, 297 articles were identified. After an initial screening for relevance and a full-text review emphasizing methodological rigor and clinical applicability, we selected 87, a curated collection of articles. This process prioritized recent advancements and practical applications in orthodontics, resulting in a robust and focused compilation of research.

## B. Literature Classification

Orthodontic treatment generally has two main phases: initial displacement caused by force application and subsequent bone remodeling. Most studies focus on the first phase. For this review, we classified articles based on their force application methods in simulations, relevant to both phases (see Figure 3).

*Activation (Loading Techniques)*

After preparing anatomical models, forces are applied using one of three techniques: (A) linear or angular appliance activation, (B) force application directly on the appliance, or (C) force application directly on the tooth (see Figure 3 (I)).

*Results (Analysis Metrics)*

The second part of simulations analyzes various parameters: initial displacement (ID), stress exerted on the appliance (SA), the force systems (FS) generated by appliance activations, stress experienced by the teeth (ST), the center of resistance (CR), stress on the bone (SB), and stress on the periodontal ligament (SPDL) (see Figure 3 (II results)). Not all studies use all these parameters for result validation.

The careful selection of loading techniques and analysis metrics is crucial for accurately simulating the biomechanical environment in orthodontic treatments, ensuring that our theoretical models closely mirror the real-world clinical scenarios

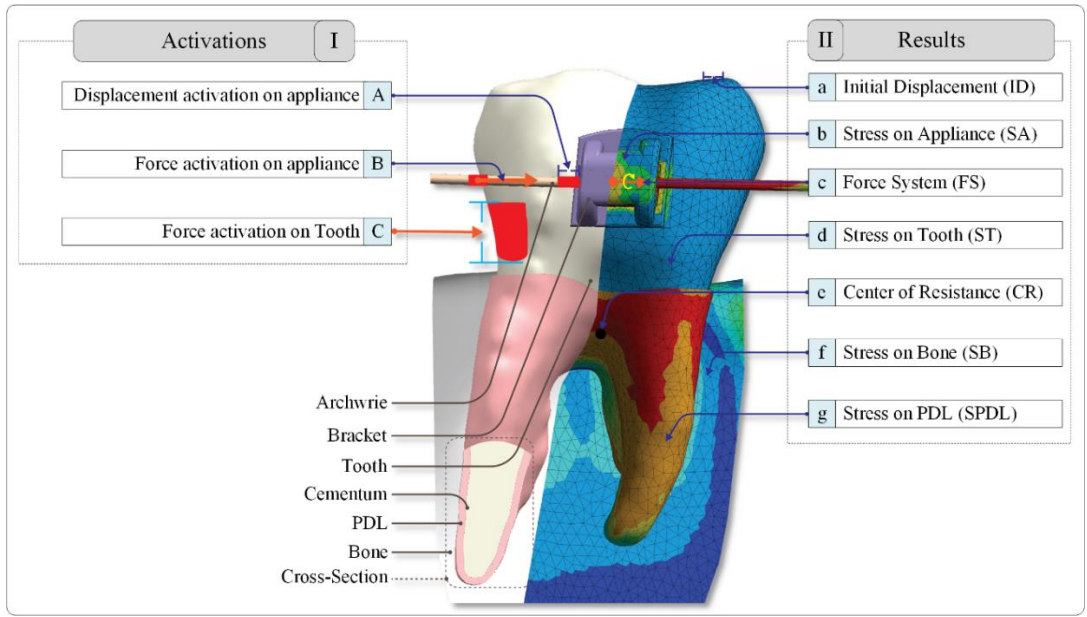

Figure 3. A detailed visualization of the simulation process for force application in fixed orthodontics, demonstrating the computational approach to analyze biomechanical interactions and stress distribution in orthodontic treatment.

Table I. Summary of the literature based on force loading techniques for different clinical scenarios

| Activations | | | Results | Models used | Clinical Scenarios | Major Findings | References |
|---|---|---|---|---|---|---|---|
| (A) | (B) | (C) | ID, SA, FS, ST, CR, SB, SPDL | | | | |
| (A) Displacement activation on appliance | | | ID: Initial Displacement SA: Stress on Appliance SB: Stress on Bone | Maxillary TPBC | Palatal expansion | The effectiveness of maxillary expansion treatments is significantly influenced by both the type of expander and suture interdigitation | [17] |
| | | | | Maxillary Skull | Palatal expansion | The study reveals how various surgical techniques and bone-borne forces impact the biomechanics of rapid palatal expansion | [18] |
| | | | | Maxillary skull | Palatal expansion | Bicortical mini-implant anchorage enhances stability and effectiveness in bone-borne palatal expansion | [19] |
| | | | | Single tooth | Bracket position | bracket's position on the tooth and its prescription significantly influence crown tip and apex displacement during orthodontic torque applications. | [20] |
| | | | ID: Initial Displacement SPDL: Stress on PDL | Partial TPBC | Palatal expansion | Screw position in the Hyrax expander affects tooth movement and stress, with more occlusal and anterior placements increasing effects. | [21] |
| | | | | Partial dentition | Space closure | Gum metal springs are more effective but slower than titanium molybdenum alloy springs in orthodontic space closure. | [22] |
| | | | | Single tooth TPBC | Tooth Tipping | Rectangular attachments in invisible orthodontics notably enhance the movement control of mandibular central incisors | [23] |
| | | | ID: Initial Displacement SB: Stress on Bone | Maxillary TPBC | Palatal expansion | Vertical vectors of anchor screws in C-expanders markedly influence palatal expansion patterns. | [24] |
| | | | | Maxillary TPBC | Palatal expansion | significant impact of bone-borne and tooth-borne expanders on stress and strain in the midpalatal suture during orthodontic treatment | [25] |
| | | | | Partial dentition | Palatal expansion | Hyrax expander screw position crucially affects maxilla's stress distribution during expansion. | [26] |
| | | | ID: Initial Displacement SA: Stress on Appliance FS: Force System | Partial dentition | Tooth extrusion | Glass fiber reinforced shape memory polyurethane enhances mechanical properties and shape recovery in orthodontics. | [27] |
| | | | | Maxillary TPBC | Space closure | Attachments in plastic aligners significantly improve diastema closure in maxillary dentition, enhancing bodily tooth movement. | [28] |
| | | | | Maxillary TPBC | Palatal expansion | The study highlights how specific attachment shapes on clear aligners significantly impact central incisor movement control. | [29] |
| (B) Force activation on appliance | | | ID: Initial Displacement SPDL: Stress on PDL | Single tooth | Tooth movement | The study introduces a new finite element analysis protocol for more accurate simulation of orthodontic tooth movement. | [30] |
| | | | | Partial TPBC | En-masse movement | Simplifying interfaces in finite element orthodontic models maintains simulation accuracy. | [31] |
| | | | | Partial TPBC | Bucco-lingual movement | Different orthodontic loads impact root resorption, correlating with periodontal ligament stimulation. | [32] |
| | | | | Maxillary TPBC | Palatally impacted canine | Traction of palatally impacted canines creates varying stress on adjacent teeth, influenced by appliance design and force direction. | [33] |
| | | | | Mandibular TPBC | Mini-screw | mandibular anterior intrusion using miniscrews for skeletal anchorage is an effective method to correct deep overbite | [34] |
| | | | | Single tooth | Molar uprighting | Effective deepbite correction with miniscrews and successful molar uprighting for implant sites. | [35] |

| | | | | | |
|---|---|---|---|---|---|
| | | Single tooth TPBC | Tooth intrusion | labial orthodontic systems experience higher torque loss than lingual systems during upper incisor intrusion | [36] |
| | | Partial dentition | Distalization | the ramal plate causes more distal and extrusive displacement of posterior teeth than miniscrews during mandibular arch distalization | [37] |
| | | Partial dentition | Space closure | the clearance of archwire size in the bracket slot significantly impacts the control of posterior tooth movement during space closure | [38] |
| | *ID: Initial Displacement* | Full Dentition | Space closure | Decreasing archwire size in miniscrew sliding mechanics leads to lingual incisor tipping and extrusion. | [39] |
| | | Maxillary TPBC | Lingual orthodontics | Transverse bowing increases with lever arm height; vertical bowing varies based on lever arm position relative to the canine. | [40] |
| | | Maxillary TPBC | Space closure | Power arms at the canines' distal side cause more lingual and labial crown tipping and reduce intercanine width in lingual orthodontics. | [41] |
| | | Maxillary TPBC | En-masse movement | Lever arm placement between lateral incisor and canine with 8 mm miniscrew height is preferred for en masse retraction in lingual treatment. | [42] |
| | | Maxillary dentition | Distalization | Mandibular dentition movement is controlled by force angulation and archwire elasticity in total distalization. | [43] |
| | | Maxillary dentition | Mesialization | Force angulation significantly influences tooth movement during maxillary dentition mesialization. | [44] |
| | *ID: Initial Displacement*<br>*CR: Center of Resistance*<br>*SPDL: Stress on PDL* | Maxillary TPBC | Anterior retraction | Identified how force angulation affects the biomechanical process of maxillary dentition mesialization | [45] |
| | | Mandibular dentition | Mesialization | Developed a method for predicting the optimal bending angles of a running loop for the effective protraction of a molar using FE and beam theory. | [46] |
| | | Mandibular dentition | Mesialization | The study identified optimal bending angles for mesial molar translation using a running loop with indirect skeletal anchorage. | [47] |
| | | Mandibular dentition | Mesialization | created a new finite element model for complete mandibular dentition mesialization. | [48] |
| *(C)*<br>*Force activation on tooth* | *ID: initial Displacement*<br>*SPDL: Stress on PDL* | Maxillary TPBC | PDL Materials | The paper enhances understanding of the biomechanical properties of the periodontal ligament (PDL) in orthodontic tooth movement | [49] |
| | | Partial TPBC | Occlusal forces | Advanced the understanding of orthodontic effects on mandibular incisors in patients with alveolar bone loss | [50] |
| | | Single tooth | Molar up-righting | The study provides insights into the biomechanics of molar uprighting with selective osteotomy and corticotomy | [51] |
| | *ID: Initial Displacement*<br>*CR: Center of Resistance* | Single tooth | Buco-palatal movement | Explored the influence of tooth dimensions on initial mobility and resistance center positioning, using numerical models from X-rays and casts. | [52] |
| | | Single tooth | Tooth movement | Presented a 3D nonlinear model for predicting tooth movement, integrating force systems and root morphology. | [53] |
| | | Mandibular TPBC | Force application conditions | Identified the 3D center of resistance in various configurations of the mandibular dentition using finite-element analysis | [54] |
| | *ID: Initial Displacement*<br>*SB: Stress on Bone* | Maxillary Skull | Palatal expansion | Presented a FE method to analyze stress and displacement patterns in the craniofacial skeleton during rapid maxillary expansion. | [55] |
| | | Maxillary dentition | Palatal expansion | The paper evaluates the effects of different parameters in miniscrew-assisted rapid palatal expansion using 3D finite element analysis. | [56] |

## 3. Category I: Displacement Activation in Orthodontic Appliances [A]

Orthodontic treatment aims to correct dental misalignment through appliance-based force application. The effectiveness of this treatment relies on the nature of the orthodontic force applied. Various studies use numerical simulations, especially focusing on inducing linear or angular displacement in orthodontic appliances for analysis. These studies can be sorted based on appliance type and evaluation parameters.

This method, aimed at simulating orthodontic forces, involves displacing virtual brackets to understand the load and movement on the dentition pre and post-treatment. Node displacements of the archwire are obtained through this process, playing a crucial role in force computation [57].

The study by Yongqing Cai investigates the interaction between brackets and various archwires, focusing on geometry and friction coefficients. This research is pivotal in understanding how these variables affect tooth rotation, inclination, and translation [58].

Fathallah et al.'s work incorporates the stiffness behavior of NiTi archwires in FE analysis, providing insights into the interaction between orthodontic appliances and bone remodeling. The study tracks bone changes as initial forces decrease over time, crucial for understanding long-term treatment dynamics [57, 59].

Girsa et al. explored the effects of pre-stressed T-loops in lingual orthodontics, while Chacko et al. compared the forces and moments generated by closed helical loops and T-loops. Tamaya et al.'s study on β-titanium alloy springs for retraction adds to this body of research, underlining the significance of loop mechanics in effective orthodontic treatment [22, 59, 60].

The literature covers a range of studies on maxillary expanders. Mieszala et al. examined polymer polyether ether ketone arches, while Gómez et al., Fernandez et al., and Choi et al. focused on the optimization of expansion screw placement. Hartono et al. and Guerrero-Vargas et al. provided comparative analyses of stress and displacement during activation. These studies collectively contribute to the understanding of maxillary expander design and its biomechanical effects [17, 19, 21, 24, 26, 29, 61-64].

Shen et al. evaluated the biomechanical impact of various mandibular expanders, focusing on their transverse displacement and efficiency. This research is integral in understanding the dynamics of mandibular expansion and its clinical implications [65].

The study of displacement activation in orthodontic appliances has significantly advanced our comprehension of orthodontic force dynamics. Focused research on various appliances, materials, and their mechanical properties has been instrumental in enhancing treatment efficacy. These investigations delve into the complexities of orthodontic force application, aiming to optimize appliance design and improve clinical outcomes, thereby contributing to more efficient and patient-specific orthodontic care.

## 4. Category II: Force Activation on Appliance [B]

Another method of loading the appliance in numerical simulation-based orthodontics involves applying force to the appliance surface in the same magnitude as used in clinical treatment. Myriad attempts have been made to compute the expected outcome of orthodontic treatment using FE to investigate the biomechanics of orthodontic appliances.

A key area of research has been the interaction between orthodontic brackets and archwires. Studies, such as those by Harikrishnan et al. and Bouton et al., have focused on understanding how forces applied to brackets affect tooth movement [30, 66]. Liu et al., emphasized the importance of detailed modeling in accurately predicting stress locations [31]. Additionally, Zhong et al., investigated the relationship between orthodontic loading and root resorption, and Maheshwari et al., examined the impact of bracket position on the stress experienced by the periodontal ligament [32, 67].

The biomechanical aspects of maxillary expansion have been scrutinized in several studies. Ulusoy et al., investigated the forces generated by appliances for unilateral maxillary expansion [68]. Comparative studies by Park et al. and Shi et al., have provided insights into different expansion techniques and their impacts on the maxilla [69, 70].

Research into the biomechanics of mandibular expansion, a critical aspect for addressing transversal discrepancies, has seen contributions from studies like Zhao et al., and Ajmera et al., [15, 71]. These studies have explored the impact of force application techniques in mandibular expansion.

The role of miniscrews in orthodontics, especially in skeletal anchorage, has been the focus of various studies. The biomechanics of miniscrews, explored by Castillo et al., and Ye et al., have provided significant insights into their effective application in clinical orthodontics [34, 72].

The mechanics of looping archwires, crucial in procedures like tooth extraction and space closure, have been extensively analyzed. Research in this area, including studies by Buyuk et al., and Cai et al., has enhanced our understanding of the biomechanics involved in these orthodontic methods [73, 74].

Headgear, traditionally used for molar distalization, has been studied to understand its biomechanical effects. Research by Feizbakhsh et al., and Alosman et al., has provided valuable insights into the stress distribution and efficacy of different headgear designs [75, 76].

The Gerber space regainer, known for its efficiency in tooth movement, has been analyzed for its biomechanical properties. Hakim et al., investigated the stress and displacement distribution when using these appliances, offering insights into their effectiveness in clinical orthodontics [77].

The comprehensive examination of force activation on orthodontic appliances offers profound insights into the biomechanics of orthodontic treatment. This body of research illuminates the intricate aspects of force application, from tooth movement to biological responses and appliance dynamics. The findings from these studies are crucial in shaping precise, effective, and safer orthodontic treatment methodologies, ensuring a more tailored approach to patient care in the field of orthodontics.

## 5. Category III: Force Activation on The Tooth [C]

Another method of loading the appliance in numerical simulation-based orthodontics involves applying force to the appliance surface in the same magnitude as used in clinical treatment. Myriad attempts have been made to compute the expected outcome of orthodontic treatment using FE to investigate the biomechanics of orthodontic appliances.

The application of direct force and moment to teeth, particularly mesially tipped mandibular molars, has been a focal area of study. She et al. aimed to evaluate the immediate mechanical impact of different corticotomy and osteotomy techniques in terms of molar uprighting [51]. Their research indicated that combining mesial and distal osteotomies with horizontal and circumferential corticotomies at the root apex level significantly enhanced tooth movement and affected the strain on the PDL differently.

Investigating the displacement and strains within the PDL of maxillary front teeth with periodontal compromise has been critical. Cortez et al. focused on evaluating the effects of different force systems applied during the early stages of space closure and en-masse retraction [78]. Their study suggested the potential suitability of cantilever mechanics for extruded front teeth with periodontal issues.

The development of mathematical models to predict orthodontic tooth movement and resistance axes has been another significant advancement. Savignano et al., developed a model based on tooth morphology that accounted for variations in the Moment to Force (M/F) ratio [53]. Additionally, Priyadarshini et al. applied a force of approximately 190 N to simulate maxillary expansion [55].

Exploring tooth rotation and its impact has been essential, especially concerning maxillary TPBC models. Păcurar et al., examined the effects of forces ranging from 1-4N, finding that forces above a certain threshold could induce dangerously high cumulative stresses [79]. Similarly, Wu et al. investigated the optimal rotational moment for teeth, uncovering varying results across different teeth [80].

The influence of occlusal forces on teeth, PDL, and bone has been a key area of research. Zeng et al., focused on the effect of occlusal loads in the mandibular anterior area, especially in patients with varying types of alveolar bone loss before orthodontic treatment [50]. Liu et al., reported that varied levels of tooth penetration into the maxillary sinus floor significantly affected the orthodontic force system aimed at achieving bodily movement [81]. Furthermore, Vilela et al., evaluated tooth mobility under physiological chewing forces and recommended specific types of splints for stabilizing avulsed teeth based on their research findings [82].

The integration of these research findings into clinical orthodontics has been vital. Understanding the direct force application on teeth, the biomechanical effects of various force systems, and the resultant stress and strain on dental and periodontal structures has greatly enhanced the precision and effectiveness of orthodontic treatments. The studies referenced here provide a comprehensive overview of the complexities and nuances involved in force activation on the tooth, contributing significantly to the field of orthodontics.

In summary, the comprehensive analysis of force activation on teeth reveals a multifaceted interplay of biomechanical forces and orthodontic procedures. The studies highlight the importance of precise force application and its impact on tooth movement, stress distribution, and the periodontal ligament. These insights are crucial in refining orthodontic treatments, ensuring they are not only effective but also minimally invasive. The findings underscore the necessity of an individualized approach, considering the unique biomechanical environment of each patient, to optimize orthodontic outcomes and reduce potential adverse effects. This approach aligns with the overarching goal of advancing orthodontic science for better patient care.

# 6. Force Loading Techniques in Invisible Orthodontics

Invisible orthodontics, primarily involving the use of aligners, has transformed the landscape of orthodontic treatment. These clear, removable braces apply force to the teeth to correct malocclusion, offering aesthetic appeal, hygienic benefits, and a minimally invasive approach. The treatment's efficacy relies on the discrepancy between the aligner's shape and the current positioning of the teeth, with this mismatch creating the necessary force for tooth movement. Advanced simulations using patient-specific dental scans and computer modeling play a crucial role in creating a series of aligners for progressive teeth alignment, as seen in Figure 4.

**Loading Activations for Analyzing Aligner Mechanics:** In the study of aligner mechanics, the challenge lies in determining where force and moment are applied, as aligners cover the entire crown of a tooth. Ma and Li analyzed optimal displacement for patients with varying degrees of inclined anterior teeth, finding an optimal displacement range of 0.1-0.18mm [83]. Rossini et al., conducted a comprehensive investigation into the biomechanical effects of incisor intrusion in open-bite cases [84]. Liu et al. proposed a novel approach using an anterior miniscrew and a plastic aligner for anterior teeth retraction [85]. Savignano et al., used an aligner with a palatal attachment to correct an anterior open bite by extruding anterior teeth, suggesting that a rectangular palatal attachment could enhance appliance efficacy [86]. Jiang et al., assessed the behavior and stress distributions of teeth under varying degrees of intrusion and retraction using clear aligners, observing tendencies for lingual root movement [83]. Hee Seo et al.'s study investigated the impact of aligner thickness on the tooth's center of rotation and stress distribution within the PDL, concluding that thicker aligners exert a slightly higher load [87]. Barone et al., explored the influence of various auxiliary elements on orthodontic tipping movement, emphasizing the importance of careful selection of these elements to optimize treatment outcomes [88].

**Loading Activations for Analyzing Attachment Mechanics:** The use of attachments in orthodontics, particularly in aligner-based treatments, has significantly enhanced control over tooth movement. Goto et al., conducted a mechanical analysis of space closure using thermoplastic aligners, examining the effects of attachments on force, moment, and stress [89]. Yokoi et al. analyzed aligner performance with optimal attachment designs for space closure in anterior teeth, finding that optimal attachments generate higher force and moment values[28]. Hong et al. introduced an overhanging attachment design for inducing bodily movement, analyzing stress distribution for minimal displacement [90]. Cortona et al. focused on the rotational movement of a second premolar in the lower arch, exploring various attachment configurations with clear aligners [91]. Their study indicates that single attachment models demonstrated high efficiency. Lastly, Gomez et al., investigated force systems and displacement patterns in orthodontic tooth movement, comparing cases with and without composite attachments, highlighting the significant impact of these attachments on tooth movement [92].

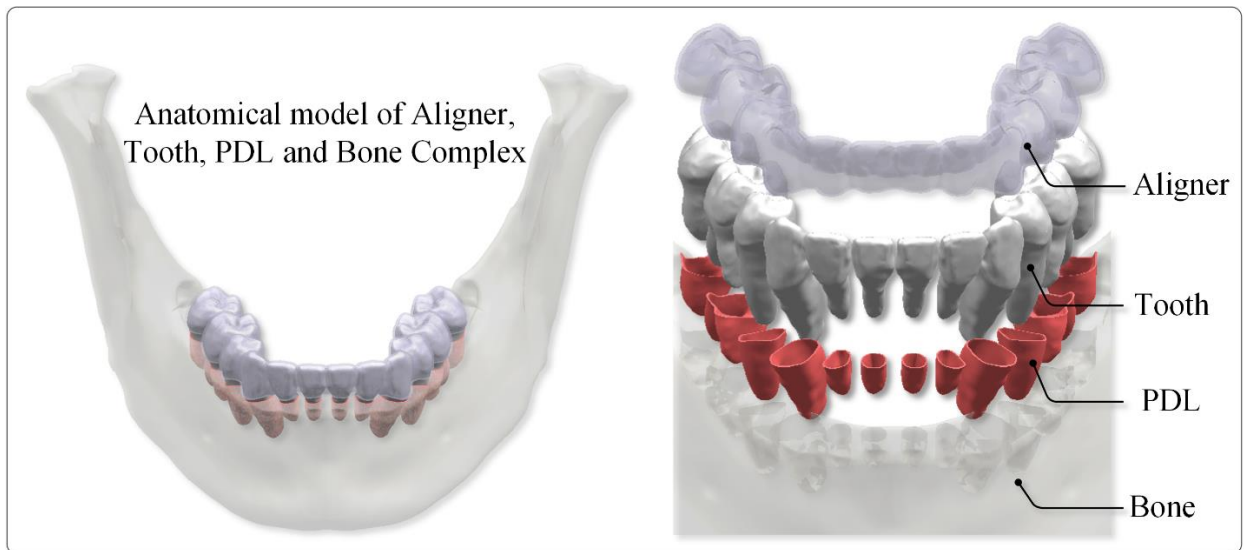

Figure 4. Conceptual overview of force simulations in invisible orthodontics, highlighting the role of aligner discrepancy and force application in tooth alignment correction

In summary, the evolution of invisible orthodontics through advanced force loading techniques and attachment mechanics represents a significant stride in orthodontic treatment, offering effective, discreet, and patient-friendly options. These advancements underscore the importance of precision and customization in treatment planning, ensuring optimal outcomes for patients seeking orthodontic correction.

## 7. Orthodontic Tooth Movement using FE

Orthodontics research has increasingly focused on using FE analysis to investigate tooth movement, particularly emphasizing initial displacement studies. However, a more comprehensive understanding of orthodontic tooth movement, especially in the context of long-term simulations incorporating alveolar bone remodeling, remains a relatively underexplored area within FE analysis. This gap in research is beginning to be addressed, with several key studies emerging to shed light on this complex aspect of orthodontics.

A few pivotal studies have concentrated on the movement of a single tooth within FE simulations [93, 94]. These studies have provided valuable insights into the initial stages of tooth movement, laying the groundwork for more intricate multi-tooth movement simulations.

Expanding the scope, other researchers have delved into the dynamics of multiple teeth movements. Notable studies in this area have explored various aspects of tooth movement, ranging from the initial displacement to more complex scenarios involving multiple teeth and their interactions with orthodontic appliances [95-97].

Significant progress has been made in understanding the interplay between brackets and archwire within a multi-tooth context[98]. This research has been instrumental in advancing our knowledge of the mechanics involved in orthodontic appliances and their effect on tooth movement.

Contributions from Bourauel et al. and Schneider et al. have been particularly influential, employing theories of bone remodeling to replicate orthodontic tooth movements accurately [93, 94]. These studies have opened new avenues in understanding the long-term effects of orthodontic treatment.

Furthering this research, Kojima and Fukui and Kojima et al. have utilized FE simulation to examine the impact of force directions on tooth movement [16, 96]. Their work has been crucial in understanding how different force applications influence the movement, especially during the closure of extraction spaces.

Hamanaka et al. and Wang et al. have introduced novel methodologies to simulate long-term orthodontic tooth movement [95, 98]. These methodologies, incorporating advanced mechanics and patient-specific models, have provided a more accurate representation of clinical situations.

In a recent study, Tamaya et al. utilized a computational model to examine long-term tooth movement [22]. Their study incorporated a double loop spring made of Gum Metal or titanium molybdenum alloy, simulating the elastic deformation of the PDL and iteratively adjusting the alveolar socket position. This approach represents a significant advancement in simulating tooth movement over extended periods, offering insights into the long-term effects of orthodontic treatments.

These diverse studies collectively mark significant progress in the field of orthodontic research, especially in understanding long-term tooth movement and alveolar bone remodeling using FE analysis. However, the complexity of this phenomenon calls for further research to deepen our understanding and refine our methodologies. By continuing to explore and innovate in this area, the orthodontics field can advance towards more accurate, patient-specific treatments, enhancing the efficacy and predictability of orthodontic care.

## 8. Discussion

In orthodontics, simulating orthodontic forces is essential in order to accurately predict orthodontic tooth movement and provide practitioners with a detailed reference point for treatment planning. A number of studies have reported detailed analyses of various clinical scenarios, with FE analysis proving to be the most efficient method for simulating orthodontic forces. FE is the leading non-destructive mechanical analysis method since it provides an accurate representation of tissue morphology, material properties, loads, and boundary conditions. However, a vast majority of FE studies focus on traditional fixed appliances and only a few studies address invisible orthodontic appliances. Invisible orthodontics, however, is growing rapidly due to its hygienic, aesthetic, and comfortable characteristics [31]. Based on the review of the existing literature, it is evident that existing approaches are fragmented and piecewise, providing a comprehensive view of only a single aspect of the entire process at a time.

Another critical point of existing studies is that they only discuss the instant response of the tooth, PDL, and bone ligament when force is applied. In orthodontic treatment, tooth movement is repeated in two phases at each appointment for the patient. During the first phase of orthodontic treatment, the instant orthodontic force causes the PDL to deform, resulting in an "initial displacement" phenomenon. Bone remodeling is triggered by stress distribution during the second phase. Periodontal tissues undergo biomechanical and histological modifications due to repeated iterations of mechanical and biological processes. The majority of studies focus on the instant force, while previous studies have not taken into account the change in force over a prolonged period of time.

Furthermore, the computation time involved in the simulation process is one of the key elements of the existing study. The literature suggests that the current FE method is reliable, robust, and capable of predicting the response to clinical scenarios with high precision. However, preprocessing and core processing requires a considerable amount of time. This is because it is necessary to construct anatomically realistic models from CT/CBCT images and appliances. Furthermore, the more detailed a model, the greater the amount of computation time required. It has also been suggested in some studies that simpler models can be used to reduce the computation and preprocessing times; however, more detailed models can provide more accurate and comprehensive results, which is a crucial requirement during clinical planning [31, 99].

In summary, to simulate the force required for instant response analysis, a continuous simulation process is required that begins with the appliance, proceeds to the tooth, then to the PDL, then to the bone complex. Additionally, to have a more reliable prediction of the future treatment, not only the instant force but the change in force over a prolonged period of time needs to be considered; therefore, tooth movement simulation will need to be investigated. Furthermore, more advanced simulation with automatizing 3D model construction and even the simulation process using deep learning techniques will be highly beneficial to the field of orthodontics which can be time efficient and more accurate. Despite the fact that automation of the orthodontic simulation process has not yet been fully addressed [100] in the literature, there have been some attempts that if successful can make this procedure reasonably efficient [101, 102].

In conclusion, this review underscores the significant progress and ongoing challenges in computational orthodontic force simulation. The insights from various simulation models have been pivotal in deepening our understanding of orthodontic mechanics. However, there is an ongoing need for models that better integrate long-term biological responses and patient-specific factors. The future of orthodontic simulations is likely to involve a blend of advanced computational methods and comprehensive biological data, leading to more individualized and effective treatment strategies. Such advancements in simulation technology are set to profoundly influence clinical orthodontic practices, ultimately enhancing patient outcomes.

## 9. Conclusion

In this review, we summarize the complexity of force-loading simulations using the FE method, their limitations, and potential in the area of fixed and invisible orthodontics. Simulations of forces are commonly carried out using FE analysis in the literature. Currently, most studies are conducted on fixed orthodontics, while few focuses on invisible orthodontics. The literature shows that available orthodontic force simulations only consider the instantaneous response to the applied force and are piecewise and fragmented, covering only a small portion of the overall process. A number of modifications have also been made to the anatomical models, both in terms of morphology and material, which could affect the final results. Future studies should concentrate on integrating advanced computational methods, including deep learning, to enhance predictive accuracy. This advancement promises to refine orthodontic treatment strategies, ultimately leading to improved patient-specific care. The continuous evolution in simulation methodologies is crucial for achieving superior clinical outcomes in orthodontics.

**Conflicts of Interest:**

The authors declare no conflict of interest.

**Acknowledgment:**

This work was supported by the National Natural Science Foundation of China (U2013205, 12426305, 62073309), partially supported by SJTU Talent Program (WH220302002), Guangdong Basic and Applied Basic Research Foundation under Grant 2022B1515020042, and Shenzhen Science and Technology Program under Grants JCYJ20220818101603008.

**Ethical Statement:**

None.

**References**

1. Proffit, W.R., H.W. Fields Jr, and D.M. Sarver, Contemporary orthodontics. 2006: Elsevier Health Sciences.
2. Jang, A., et al., A force on the crown and tug of war in the periodontal complex. Journal of dental research, 2018. **97**(3): p. 241-250.
3. Graves, D.T., et al., Osteoimmunology in the oral cavity (periodontal disease, lesions of endodontic origin, and orthodontic tooth movement), in Osteoimmunology. 2016, Elsevier. p. 325-344.
4. Li, Y., et al., Biomechanical and biological responses of periodontium in orthodontic tooth movement: up-date in a new decade. International journal of oral science, 2021. **13**(1): p. 20.
5. Bichu, Y.M., et al., Applications of artificial intelligence and machine learning in orthodontics: a scoping review. Progress in orthodontics, 2021. **22**(1): p. 1-11.
6. Shi, B. and H. Huang, Computational technology for nasal cartilage-related clinical research and application. International Journal of Oral Science, 2020. **12**(1): p. 21.
7. Farah, J., R.G. Craig, and D.L. Sikarskie, Photoelastic and finite element stress analysis of a restored axisymmetric first molar. Journal of biomechanics, 1973. **6**(5): p. 511-520.
8. Duanmu, Z., et al., Development of a biomechanical model for dynamic occlusal stress analysis. International Journal of Oral Science, 2021. **13**(1): p. 29.
9. Cattaneo, P., M. Dalstra, and B. Melsen, The finite element method: a tool to study orthodontic tooth movement. Journal of dental research, 2005. **84**(5): p. 428-433.
10. McCormack, S.W., et al., Inclusion of periodontal ligament fibres in mandibular finite element models leads to an increase in alveolar bone strains. PLoS one, 2017. **12**(11): p. e0188707.
11. Ammar, H.H., et al., Three-dimensional modeling and finite element analysis in treatment planning for orthodontic tooth movement. American Journal of Orthodontics and Dentofacial Orthopedics, 2011. **139**(1): p. e59-e71.
12. Park, Y., et al., Deep Learning–Based Prediction of the 3D Postorthodontic Facial Changes. Journal of Dental Research, 2022. **101**(11): p. 1372-1379.
13. Wang, H., et al., Multiclass CBCT image segmentation for orthodontics with deep learning. Journal of Dental Research, 2021. **100**(9): p. 943-949.
14. Kim, W.-H., et al., Optimal position of attachment for removable thermoplastic aligner on the lower canine using finite element analysis. Materials, 2020. **13**(15): p. 3369.
15. Zhao, B., et al., A pilot study of mandibular expansion in combination with a fixed-appliance for increasing the effective space of the mandibular arch: Finite element analysis and three-dimensional cone-beam computed tomography. Medicine, 2021. **100**(8).
16. Kojima, Y., J. Kawamura, and H. Fukui, Finite element analysis of the effect of force directions on tooth movement in extraction space closure with miniscrew sliding mechanics. American journal of orthodontics and dentofacial orthopedics, 2012. **142**(4): p. 501-508.
17. Guerrero-Vargas, J., et al., Influence of interdigitation and expander type in the mechanical response of the midpalatal suture during maxillary expansion. Computer methods and programs in biomedicine, 2019. **176**: p. 195-209.
18. Möhlhenrich, S., et al., Simulation of three surgical techniques combined with two different bone-borne forces for surgically assisted rapid palatal expansion of the maxillofacial complex: a finite element analysis. International Journal of Oral and Maxillofacial Surgery, 2017. **46**(10): p. 1306-1314.
19. Lee, R.J., W. Moon, and C. Hong, Effects of monocortical and bicortical mini-implant anchorage on bone-borne palatal expansion using finite element analysis. American Journal of Orthodontics and Dentofacial Orthopedics, 2017. **151**(5): p. 887-897.
20. Papageorgiou, S.N., et al., Torque differences according to tooth morphology and bracket placement: a finite element study. European journal of orthodontics, 2017. **39**(4): p. 411-418.
21. Fernandes, L.C., et al., Influence of the hyrax expander screw position on displacement and stress distribution in teeth: A study with finite elements. American Journal of Orthodontics and Dentofacial Orthopedics, 2021. **160**(2): p. 266-275.
22. Tamaya, N., J. Kawamura, and Y. Yanagi, Tooth Movement Efficacy of Retraction Spring Made of a New Low Elastic Modulus Material, Gum Metal, Evaluated by the Finite Element Method. Materials, 2021. **14**(11): p. 2934.
23. Pan, S., et al., Effect of Attachment on Movement Control of the Central Incisor Using Invisible Orthodontics: In-Silico Finite Element Analysis. Journal of Shanghai Jiaotong University (Science), 2021. **26**(3): p. 383-390.
24. Choi, J.-Y., et al., Finite element analysis of C-expanders with different vertical vectors of anchor screws. American Journal of Orthodontics and Dentofacial Orthopedics, 2021. **159**(6): p. 799-807.
25. Trojan, L.C., et al., Stresses and strains analysis using different palatal expander appliances in upper jaw and midpalatal suture. Artificial Organs, 2017. **6**(41): p. E41-E51.
26. Fernandes, L.C., et al., Influence of the hyrax expander screw position on stress distribution in the maxilla: A study with finite elements. American Journal of Orthodontics and Dentofacial Orthopedics, 2019. **155**(1): p. 80-87.
27. Liu, Y.-f., et al., Thermo-mechanical properties of glass fiber reinforced shape memory polyurethane for orthodontic application. Journal of Materials Science: Materials in Medicine, 2018. **29**(9): p. 1-11.
28. Yokoi, Y., et al., Effects of attachment of plastic aligner in closing of diastema of maxillary dentition by finite element method. Journal of Healthcare Engineering, 2019. **2019**.
29. Mieszala, C., et al., Digital Design of Different Transpalatal Arches Made of Polyether Ether Ketone (PEEK) and Determination of the Force Systems. Applied Sciences, 2022. **12**(3): p. 1590.
30. Bouton, A., et al., New finite element study protocol: clinical simulation of orthodontic tooth movement. International Orthodontics, 2017. **15**(2): p. 165-179.
31. Liu, Y., F. Jiang, and J. Chen, Can interfaces at bracket-wire and between teeth in multi-teeth finite element model be simplified? International Journal for Numerical Methods in Biomedical Engineering, 2019. **35**(3): p. e3169.
32. Zhong, J., et al., In vivo effects of different orthodontic loading on root resorption and correlation with mechanobiological stimulus in periodontal ligament. Journal of the Royal Society Interface, 2019. **16**(154): p. 20190108.
33. Zeno, K.G., et al., Finite element analysis of stresses on adjacent teeth during the traction of palatally impacted canines. The Angle Orthodontist, 2019. **89**(3): p. 418-425.
34. del Castillo McGrath, M.G., et al., Mandibular anterior intrusion using miniscrews for skeletal anchorage: A 3-dimensional finite element analysis. American Journal of Orthodontics and Dentofacial Orthopedics, 2018. **154**(4): p. 469-476.
35. Zhou, J., et al., Selective osteotomy-assisted molar uprighting and simultaneous ridge augmentation for implant site development. American Journal of Orthodontics and Dentofacial Orthopedics, 2019. **156**(6): p. 846-857.
36. Pol, T.R., et al., Torque control during intrusion on upper central incisor in labial and lingual bracket system-a 3D finite element study.

Journal of Clinical and Experimental Dentistry, 2018. **10**(1): p. e20.
37. Kim, Y.B., et al., Displacement of mandibular dentition during total arch distalization according to locations and types of TSAD s: 3D Finite element analysis. Orthodontics & craniofacial research, 2019. **22**(1): p. 46-52.
38. Ahmed, N., et al., Effect of bracket slot and archwire dimension on posterior tooth movement in sliding mechanics: A Three-dimensional finite element analysis. Cureus, 2019. **11**(9).
39. Kawamura, J. and N. Tamaya, A finite element analysis of the effects of archwire size on orthodontic tooth movement in extraction space closure with miniscrew sliding mechanics. Progress in Orthodontics, 2019. **20**(1): p. 1-6.
40. Mattu, N., A.M. Virupaksha, and A. Belludi, Comparative study of effect of different lever arm positions and lengths on transverse and vertical bowing in lingual orthodontics–An FEM study. International Orthodontics, 2021. **19**(2): p. 281-290.
41. Feng, Y., et al., Finite element analysis of the effect of power arm locations on tooth movement in extraction space closure with miniscrew anchorage in customized lingual orthodontic treatment. American Journal of Orthodontics and Dentofacial Orthopedics, 2019. **156**(2): p. 210-219.
42. Ghannam, M. and B. Kamiloğlu, Effects of Skeletally Supported Anterior en Masse Retraction with Varied Lever Arm Lengths and Locations in Lingual Orthodontic Treatment: A 3D Finite Element Study. BioMed Research International, 2021. **2021**.
43. Chae, J.-M., et al., Biomechanical analysis for total distalization of the mandibular dentition: a finite element study. American Journal of Orthodontics and Dentofacial Orthopedics, 2019. **155**(3): p. 388-397.
44. Kawamura, J., et al., Biomechanical analysis for total mesialization of the maxillary dentition: A finite element study. American Journal of Orthodontics and Dentofacial Orthopedics, 2021. **159**(6): p. 790-798.
45. Park, J.H., et al., Palatal en-masse retraction of segmented maxillary anterior teeth: A finite element study. The korean journal of orthodontics, 2019. **49**(3): p. 188-193.
46. Ryu, W.-K., et al., Prediction of optimal bending angles of a running loop to achieve bodily protraction of a molar using the finite element method. The Korean Journal of Orthodontics, 2018. **48**(1): p. 3-10.
47. Kim, M.J., et al., A finite element analysis of the optimal bending angles in a running loop for mesial translation of a mandibular molar using indirect skeletal anchorage. Orthodontics & Craniofacial Research, 2018. **21**(1): p. 63-70.
48. Kawamura, J., et al., Biomechanical analysis for total mesialization of the mandibular dentition: A finite element study. Orthodontics & craniofacial research, 2019. **22**(4): p. 329-336.
49. Uhlir, R., et al., Biomechanical characterization of the periodontal ligament: Orthodontic tooth movement. The Angle Orthodontist, 2017. **87**(2): p. 183-192.
50. Zeng, Y., L. Xiao, and X. Yuan, Displacement and stress distribution of mandibular incisors after orthodontic treatment in the presence of alveolar bone loss under occlusal loads: A finite element analysis. American Journal of Orthodontics and Dentofacial Orthopedics, 2022. **161**(5): p. e456-e465.
51. She, X., et al., Biomechanical effect of selective osteotomy and corticotomy on orthodontic molar uprighting. American Journal of Orthodontics and Dentofacial Orthopedics, 2021. **160**(2): p. 292-301.
52. Hartmann, M., et al., Influence of tooth dimension on the initial mobility based on plaster casts and X-ray images. Journal of Orofacial Orthopedics/Fortschritte der Kieferorthopädie, 2017. **78**(4): p. 285-292.
53. Savignano, R., R.F. Viecilli, and U. Oyoyo, Three-dimensional nonlinear prediction of tooth movement from the force system and root morphology. The Angle Orthodontist, 2020. **90**(6): p. 811-822.
54. Jo, A.-R., et al., Finite-element analysis of the center of resistance of the mandibular dentition. The korean journal of orthodontics, 2017. **47**(1): p. 21-30.
55. Priyadarshini, J., et al., Stress and displacement patterns in the craniofacial skeleton with rapid maxillary expansion—a finite element method study. Progress in Orthodontics, 2017. **18**(1): p. 1-8.
56. Yoon, S., D.-Y. Lee, and S.-K. Jung, Influence of changing various parameters in miniscrew-

assisted rapid palatal expansion: a three-dimensional finite element analysis. The korean journal of orthodontics, 2019. **49**(3): p. 150-160.
57. Fathallah, A., et al., Three-dimensional coupling between orthodontic bone remodeling and superelastic behavior of a NiTi wire applied for initial alignment. Journal of Orofacial Orthopedics/Fortschritte der Kieferorthopadie, 2021. **82**(2).
58. Cai, Y., A three-dimensional finite element analysis of the effect of archwire characteristics on the self-ligating orthodontic tooth movement of the canine. Technology and Health Care, 2019. **27**(S1): p. 195-204.
59. Liu, Y.-f., et al., Thermo-mechanical properties of glass fiber reinforced shape memory polyurethane for orthodontic application. Journal of Materials Science: Materials in Medicine, 2018. **29**: p. 1-11.
60. Chacko, A., et al., Comparative assessment of the efficacy of closed helical loop and T-loop for space closure in lingual orthodontics—a finite element study. Progress in orthodontics, 2018. **19**(1): p. 1-8.
61. Gómez-Gómez, S.-L., et al., Influence of Hyrax screw position on dental movement and cortical bone: a study of finite elements. Journal of Clinical and Experimental Dentistry, 2019. **11**(12): p. e1099.
62. Hartono, N., B.M. Soegiharto, and R. Widayati, The difference of stress distribution of maxillary expansion using rapid maxillary expander (RME) and maxillary skeletal expander (MSE)—a finite element analysis. Progress in orthodontics, 2018. **19**: p. 1-10.
63. Nowak, R., et al., Comparison of tooth-and bone-borne appliances on the stress distributions and displacement patterns in the facial skeleton in surgically assisted rapid maxillary expansion—a finite element analysis (FEA) study. Materials, 2021. **14**(5): p. 1152.
64. Oliveira, P.L.E., et al., Stress and displacement of mini-implants and appliance in Mini-implant Assisted Rapid Palatal Expansion: analysis by finite element method. Dental Press Journal of Orthodontics, 2021. **26**.
65. Shen, T., et al., Efficacy of different designs of mandibular expanders: A 3-dimensional finite element study. American Journal of Orthodontics and Dentofacial Orthopedics, 2020. **157**(5): p. 641-650.
66. Harikrishnan, P. and V. Magesh, Stress distribution and deformation in six tie wings Orthodontic bracket during simulated tipping–A finite element analysis. Computer Methods and Programs in Biomedicine, 2021. **200**: p. 105835.
67. Maheshwari, R.K., et al., The effect of tooth morphology and vertical bracket positioning on resultant stress in periodontal ligament-a three dimensional finite element study. Medicine and Pharmacy Reports, 2019. **92**(3): p. 294.
68. Ulusoy, Ç. and M. Dogan, A new method for the treatment of unilateral posterior cross-bite: a three-dimensional finite element stress analysis study. Progress in Orthodontics, 2018. **19**(1): p. 1-9.
69. Park, J.H., et al., Displacement and stress distribution by different bone-borne palatal expanders with facemask: a 3-dimensional finite element analysis. American Journal of Orthodontics and Dentofacial Orthopedics, 2017. **151**(1): p. 105-117.
70. Shi, Y., C.-n. Zhu, and Z. Xie, Displacement and stress distribution of the maxilla under different surgical conditions in three typical models with bone-borne distraction: a three-dimensional finite element analysis. Journal of Orofacial Orthopedics/Fortschritte der Kieferorthopadie, 2020. **81**(6).
71. Ajmera, D.H., et al., Analysis of dentoalveolar structures with novel corticotomy-facilitated mandibular expansion: A 3-dimensional finite element study. American Journal of Orthodontics and Dentofacial Orthopedics, 2017. **151**(4): p. 767-778.
72. Ye, Y., et al., Optimization Analysis of Two-Factor Continuous Variable between Thread Depth and Pitch of Microimplant under Toque Force. Computational and Mathematical Methods in Medicine, 2022. **2022**.
73. Buyuk, S.K., M.S. Guler, and M.L. Bekci, Effect of arch wire size on orthodontic reverse closing loop and retraction force in canine tooth distalization. Journal of Orofacial Orthopedics/Fortschritte der Kieferorthopadie, 2019. **80**(1).
74. Cai, Y., Finite element analysis of archwire parameters and activation forces on the M/F ratio of vertical, L-and T-loops. BMC oral health, 2020. **20**(1): p. 1-12.
75. Feizbakhsh, M., et al., Stress distribution in maxillary first molar periodontium using straight

pull headgear with vertical and horizontal tubes: A finite element analysis. Dental Research Journal, 2017. **14**(2): p. 117.
76. Alosman, H.S., M. Bayome, and L. Vahdettin, A 3D finite element analysis of maxillary molar distalization using unilateral zygoma gear and asymmetric headgear. Orthodontics & Craniofacial Research, 2021. **24**(2): p. 261-267.
77. Hakim, M.A.A., et al., A comparative three-dimensional finite element study of two space regainers in the mixed dentition stage. European journal of dentistry, 2020. **14**(01): p. 107-114.
78. Frias Cortez, M.A., et al., Numerical and biomechanical analysis of orthodontic treatment of recovered periodontally compromised patients. Journal of Orofacial Orthopedics/Fortschritte der Kieferorthopadie, 2022. **83**(4).
79. Păcurar, M., et al., The effect of rotation upon dental structure components following orthodontic fix appliance. Medicine and Pharmacy Reports, 2019. **92**(Suppl No 3): p. S45.
80. Wu, J., et al., Numerical simulation of optimal range of rotational moment for the mandibular lateral incisor, canine and first premolar based on biomechanical responses of periodontal ligaments: A case study. Clinical Oral Investigations, 2021. **25**: p. 1569-1577.
81. Liu, X., et al., Effect of the Maxillary Sinus on Tooth Movement during Orthodontics Based on Biomechanical Responses of Periodontal Ligaments. Applied Sciences, 2022. **12**(10): p. 4990.
82. Vilela, A.B.F., et al., Splint stiffness and extension effects on a simulated avulsed permanent incisor—A patient-specific finite element analysis. Dental Traumatology, 2022. **38**(1): p. 53-61.
83. Ma, Y. and S. Li, The optimal orthodontic displacement of clear aligner for mild, moderate and severe periodontal conditions: an in vitro study in a periodontally compromised individual using the finite element model. BMC Oral Health, 2021. **21**: p. 1-8.
84. Rossini, G., et al., Incisors extrusion with clear aligners technique: a finite element analysis study. Applied Sciences, 2021. **11**(3): p. 1167.
85. Liu, L., et al., Effectiveness of an anterior mini-screw in achieving incisor intrusion and palatal root torque for anterior retraction with clear aligners: A finite element study. The Angle Orthodontist, 2021. **91**(6): p. 794-803.
86. Savignano, R., et al., Biomechanical effects of different auxiliary-aligner designs for the extrusion of an upper central incisor: a finite element analysis. Journal of Healthcare Engineering, 2019. **2019**.
87. Seo, J.-H., et al., Comparative analysis of stress in the periodontal ligament and center of rotation in the tooth after orthodontic treatment depending on clear aligner thickness—finite element analysis study. Materials, 2021. **14**(2): p. 324.
88. Barone, S., et al., Computational design and engineering of polymeric orthodontic aligners. International journal for numerical methods in biomedical engineering, 2017. **33**(8): p. e2839.
89. Goto, M., et al., A method for evaluation of the effects of attachments in aligner-type orthodontic appliance: Three-dimensional finite element analysis. Orthodontic Waves, 2017. **76**(4): p. 207-214.
90. Hong, K., et al., Efficient design of a clear aligner attachment to induce bodily tooth movement in orthodontic treatment using finite element analysis. Materials, 2021. **14**(17): p. 4926.
91. Cortona, A., et al., Clear aligner orthodontic therapy of rotated mandibular round-shaped teeth: a finite element study. The Angle Orthodontist, 2020. **90**(2): p. 247-254.
92. Gomez, J.P., et al., Initial force systems during bodily tooth movement with plastic aligners and composite attachments: A three-dimensional finite element analysis. The Angle Orthodontist, 2015. **85**(3): p. 454-460.
93. Bourauel, C., D. Vollmer, and A. Jäger, Application of bone remodeling theories in the simulation of orthodontic tooth movements. Journal of Orofacial Orthopedics/Fortschritte der Kieferorthopädie, 2000. **61**: p. 266-279.
94. Schneider, J., M. Geiger, and F.-G. Sander, Numerical experiments on long-time orthodontic tooth movement. American journal of orthodontics and dentofacial orthopedics, 2002. **121**(3): p. 257-265.
95. Hamanaka, R., et al., Numeric simulation model for long-term orthodontic tooth movement with contact boundary conditions using the finite element method. American

Journal of Orthodontics and Dentofacial Orthopedics, 2017. **152**(5): p. 601-612.
96. Kojima, Y. and H. Fukui, A finite element simulation of initial movement, orthodontic movement, and the centre of resistance of the maxillary teeth connected with an archwire. The European Journal of Orthodontics, 2014. **36**(3): p. 255-261.
97. Kawamura, J. and N. Tamaya, A finite element analysis of the effects of archwire size on orthodontic tooth movement in extraction space closure with miniscrew sliding mechanics. Progress in Orthodontics, 2019. **20**: p. 1-6.
98. Wang, C., et al., Simulation of bone remodelling in orthodontic treatment. Computer Methods in Biomechanics and Biomedical Engineering, 2014. **17**(9): p. 1042-1050.
99. Schmidt, F. and B.G. Lapatki, Effect of variable periodontal ligament thickness and its non-linear material properties on the location of a tooth's centre of resistance. Journal of Biomechanics, 2019. **94**: p. 211-218.
100. Niño-Sandoval, T.C., et al., An automatic method for skeletal patterns classification using craniomaxillary variables on a Colombian population. Forensic Science International, 2016. **261**: p. 159. e1-159. e6.
101. Jang, T.J., et al., A fully automated method for 3D individual tooth identification and segmentation in dental CBCT. arXiv preprint arXiv:2102.06060, 2021.
102. Cui, Z., et al., A fully automatic AI system for tooth and alveolar bone segmentation from cone-beam CT images. Nature communications, 2022. **13**(1): p. 1-11.